\documentclass[a4paper,oneside]{article}

\usepackage{microtype}
\usepackage[sc]{mathpazo}
\usepackage[T1]{fontenc}
\usepackage[DIV=11,BCOR=0cm,headinclude=true]{typearea}
\usepackage{amsfonts}
\usepackage{amsmath}
\usepackage{amssymb}
\usepackage{amsthm}
\usepackage{array}
\usepackage{graphicx}
\usepackage{float}
\usepackage{booktabs}
\usepackage[textfont=sl]{subfig}
\usepackage{caption}
\usepackage{mathrsfs}

\usepackage{authblk}

\usepackage[maxbibnames=100,autocite=superscript,style=custom-nature,sorting=none,defernumbers=true]{biblatex}
\newcommand{\bv}[1]{\boldsymbol{#1}}

\DeclareMathOperator*{\real}{Re}
\DeclareMathOperator*{\imag}{Im}

\newcommand{\overbar}[1]{\mkern 1.5mu\overline{\mkern-1.5mu#1\mkern-1.5mu}\mkern 1.5mu}

\author[1,2,$\dagger$]{\sc{Parry Y. Chen}}
\author[3]{\sc{Michael J. A. Smith}}
\author[3]{\sc{Ross C. McPhedran}}
\affil[1]{School of Physics and Astronomy, Raymond and Beverly Sackler Faculty of Exact Sciences, Tel Aviv University, Israel}
\affil[2]{Unit of Electro-optic Engineering, Ben-Gurion University, Israel}
\affil[3]{Centre for Ultrahigh bandwidth Devices for Optical Systems (CUDOS), School of Physics, The University of Sydney, NSW 2006, Australia}
\affil[$\dagger$]{\it{parryyu@post.bgu.ac.il}}

\title{Evaluation and regularization of generalized Eisenstein series and application to 2D cylindrical harmonic sums}
\date{\today}

\begin{document}

\maketitle

\begin{abstract} \noindent
In the study of periodic media, conditionally convergent series are frequently encountered and their regularization is crucial for applications. We derive an identity that regularizes two-dimensional generalized Eisenstein series for all Bravais lattices, yielding physically meaningful values. We also obtain explicit forms for the generalized series in terms of conventional Eisenstein series, enabling their closed-form evaluation for   important high symmetry lattices. Results are then used to obtain representations for the related cylindrical harmonic sums, which are also given for all Bravais lattices. Finally, we treat displaced lattices of high symmetry, expressing them  in terms of  origin-centered lattices via geometric multi-set identities. These identities apply to all classes of two-dimensional sums, allowing sums to be evaluated over each constituent of a unit cell that possesses multiple inclusions.
\end{abstract}

\section{Introduction}
The modeling of periodic structures often requires the evaluation of infinite sums over the point set of an underlying lattice.\autocite{borwein2013lattice} There is considerable interest in both the correct numerical evaluation and the analytical study of such sums, as they feature in a wide range of applications; from the calculation of effective material parameters to the study of band structures for a periodic composite material. In these contexts, the governing equation is the Helmholtz operator or Laplace's equation, from which many different classes of lattice sums can appear. In this paper, we examine two classes of lattice sums that arise frequently in two-dimensional structured media. These are the generalized Eisenstein series
\begin{equation}
\sigma_n^{(m)}(\tau) = \sideset{}{'}\sum_{p} \frac{e^{-im\varphi_p}}{R^n_p},
\label{eq:sigma}
\end{equation}
and the cylindrical harmonic sums
\begin{equation}
S_{l,m,n}(u;\tau,a) = \sideset{}{'}\sum_{h} \frac{J_l(K_hu)}{K_h^n}e^{im\psi_h},
\label{eq:Sdef}
\end{equation}
where $(R_p,\varphi_p)$ and $(K_h,\psi_h)$ are polar forms for the coordinates of a two-dimensional Bravais lattice over direct and reciprocal space, respectively, and $u$ is an arbitrary coordinate. The parameters $a$ and $\tau$ are constants that describe the underlying Bravais lattice, and the superscript $\prime$ denotes exclusion of singular terms, for example when $R_p = 0$. See Section \ref{sec:defn} for full definitions.

The two infinite sums are intimately linked, since the cylindrical harmonic sums \eqref{eq:Sdef} can be expressed in terms of generalized Eisenstein series \eqref{eq:sigma} via the Fourier transform, motivating their joint consideration in our paper. However, Eisenstein series also independently feature in effective medium calculations where the corresponding physical field satisfies the Laplace or Helmholtz operator. This includes calculations for the effective electrical and heat conductivity \autocite{rayleigh1892influence,perrins1979transport,mcphedran1978conductivity,mityushev1997transporta,mityushev1997transportb,mityushev2006representative,nicorovici1996analytical,godin2012effective,rylko2000transport,rogosin2009eisenstein}, effective permittivity,\autocite{mcphedran1996low,godin2013effective} effective permeability,\autocite{mityushev2002longitudinal} and effective elastic constants \autocite{helsing1994bounds} at long wavelengths. Eisenstein series can also be found in the static quasi-periodic Green's function for the Helmholtz operator.\autocite{movchan1997greens} For effective medium calculations, the $\sigma_n^{(m)}$ sums are often considered in the form $m=n$, reducing them to conventional Eisenstein series $G_n = \sigma_{n}^{(n)} $.\autocite{zagier1992introduction} Although modern computing resources can directly evaluate the majority of conventional and generalized Eisenstein series with ease, brute force methods fail for $n=2$, when these series become conditionally convergent. In these instances, the values obtained depend on the method of summation, which in turn depend on the geometry of the finite lattice chosen.

The issue of evaluating conditionally convergent sums can be managed with careful summation procedures,\autocite{rayleigh1892influence,rogosin2009eisenstein,yardleyphdthesis,weil2012elliptic,yakubovich2016closed} or special function representations.\autocite{godin2012effective,godin2013effective,rylko2000transport} However, many of these special numerical summation techniques only apply to certain lattices, and some of these methods simply mask the appearance of the conditionally convergent sums that are inherently present. In particular, the analytic and mathematically rigorous Eisenstein summation method always yields a definitive result by enforcing the order of summation,\autocite{rogosin2009eisenstein,weil2012elliptic} but includes a non-physical contribution that can be difficult to identify and exclude. In the current work, we provide a robust method to regularize generalized Eisenstein series for all $(m,2)$ orders and for all lattice geometries. This is achieved by analytically continuing from absolutely convergent sums, thereby enforcing geometric identities that must be satisfied to ensure physically meaningful results. The simple formula we obtain alleviates all problems associated with conditional convergence, and drastically simplifies the procedure for subsequent effective medium calculations over an arbitrary lattice. 

Next, we derive expressions to reduce the generalized Eisenstein series to products of the standard Eisenstein series, applicable to orders $m > n$. This enables the closed-form evaluation of many Eisenstein series over a variety of high symmetry lattices, which were only previously known numerically for a select few Bravais lattice geometries.\autocite{nicorovici1996analytical,stremler2004evaluation,glasser1974evaluation} The simple procedure is entirely formulaic, reducing to evaluations of the Dedekind $\eta$-function, for which many special values are known,\autocite{weber1908lehrbuch,zucker1977evaluation} and many more can be generated using the celebrated Chowla--Selberg formula.\autocite{chowla1949epsteins,vanderpoorten1999values,hart2015class} We thus explicitly evaluate and tabulate generalized Eisenstein series for a variety of frequently encountered lattices and orders, including conditionally convergent orders. For all other orders of the generalized Eisenstein series, rapidly convergent sums are obtained that are applicable to any lattice geometry. In the case of conditionally convergent orders, all evaluation methods produce results that conform to the order of summation used by the Eisenstein summation method, which can subsequently be regularized as described above. We utilize these methods in our analysis of cylindrical harmonic sums $S_{l,m,n}$, in particular demonstrating that regularized sums can be added, subtracted, and are well behaved for all further purposes including the representation of the absolutely convergent $S_{l,m,n}$.

The second class of sums $S_{l,m,n}$ play an important role in the long-wavelength representations of the quasi-periodic Green's function for the Helmholtz operator,\autocite{linton2010lattice,mcphedran2000lattice,chin1994greens} as well as in band structure and effective index calculations for photonic, phononic, and other periodic structures.\autocite{chin1994greens,mcphedran1996low,movchan2002asymptotic}   At present, recurrence relations are available to obtain $S_{l,m,n}$ analytically for square lattices. \autocite{nicorovici1996analytical} We have extended these relations to all Bravais lattices and used them to obtain   general expressions for $S_{l,m,n}$ in closed-form, which are applicable to all orders and lattices. When $l-n$ is even, these   expressions are polynomials in $u$ with regularized $\sigma_n^{(m)}$ as coefficients, whereas for odd $l-n$, cylindrical harmonic sums can only be expressed as an infinite series.\autocite{nicorovici1996analytical}

Finally, we consider the $\sigma_n^{(m)}$ and $S_{l,m,n}$ sums when they are evaluated over displaced point sets. At present, Eisenstein series and cylindrical harmonic sums are evaluated over point sets that are centered about the origin. For the reciprocal lattice, this corresponds to the set of all $\Gamma$ points in all Brillouin cells, and for the direct lattice, this corresponds to the origin of all unit cells. The framework we outline admits closed-form expressions for more complicated lattice geometries, such as diatomic lattices. It also permits a closed-form asymptotic analysis of band structures at points of high symmetry other than the $\Gamma$ point. Although the Eisenstein series and cylindrical harmonic sums considered here correspond to the static or quasi-static limit in frequency, the method we present is general and applies to all lattice sums in two dimensions: we derive multi-set identities to evaluate over sets comprising the high symmetry points of a Bravais lattice. Explicit representations for square and hexagonal lattices are presented and applied to \eqref{eq:sigma} and \eqref{eq:Sdef}.

The outline of this paper is as follows. In Section \ref{sec:defn} we provide important definitions, nomenclature, and representations of the lattices. Section \ref{sec:eisenstein} evaluates $\sigma_n^{(m)}$ over an origin-centered direct lattice and, in particular, Section \ref{sec:nonmod} discusses regularization for $n=2$. Section \ref{sec:gamma} evaluates $S_{l,m,n}$ over an origin-centered reciprocal lattice, using the results from Section \ref{sec:eisenstein} to provide expressions all lattices and applicable orders. Section \ref{sec:displaced} treats displaced square and hexagonal lattices and deriving multi-set identities for $\sigma_n^{(m)}$ and $S_{l,m,n}$. 

\section{Lattice definitions} 
\label{sec:defn}
We define an origin-centered 2D lattice in real space as the set of all points
\begin{equation}
\Omega = \{p_1a\bv{\hat{e}}_1 + p_2b\bv{\hat{e}}_2 | p_1,p_2 \in \mathbb{Z}\},
\label{eq:lattice}
\end{equation}
where $a$ and $b$ are the lengths of the unit cell edges, and unit vectors $\bv{\hat{e}}_1$ and $\bv{\hat{e}}_2$ their directions. Without loss of generality, $\bv{\hat{e}}_1$ may be oriented to lie along the $x$-axis. Alternatively, it is convenient to define a complex plane representation of $\Omega$, by orienting the $x$-axis and thus $\bv{\hat{e}}_1$ along the real axis and relating $b\bv{\hat{e}}_2$ to a single complex parameter,
\begin{equation}
\tau = \frac{b}{a}e^{i\phi},
\label{eq:tau}
\end{equation}
where $\phi$ is the angle between $\bv{\hat{e}}_1$ and $\bv{\hat{e}}_2$. By convention, $\tau$ is chosen to lie in the upper half of the complex plane. The lattice is thus uniquely specified by $\Omega(\tau,a)$.

The reciprocal lattice of $\Omega$ is defined by
\begin{equation}
\overbar{\Omega} = \left.\left\{\frac{2\pi}{A}(h_1b\bv{\hat{e}}'_1 + h_2a\bv{\hat{e}}'_2)\right|h_1,h_2 \in \mathbb{Z}\right\},
\label{eq:rlattice}
\end{equation}
where $A = ab\sin\varphi$ is the area of the unit cell, and $\bv{\hat{e}}'_1$ and $\bv{\hat{e}}'_2$ are the unit reciprocal lattice vectors, chosen such that $\bv{\hat{e}}_1 \cdot \bv{\hat{e}}'_2 = \bv{\hat{e}}_2 \cdot \bv{\hat{e}}'_1 = 0$. Since $\overbar{\Omega}$ is derived from $\Omega$, it may also be uniquely specified by $\overbar{\Omega}(\tau,a)$.

For the generalized Eisenstein sums \eqref{eq:sigma}, the index $p = (p_1,p_2)$ runs over all points of the lattice $\Omega(\tau,1)$, with each point given in polar coordinates $(R_p,\varphi_p)$. For convenience, $\sigma_n^{(m)}(\tau)$ is defined only for $a = 1$, as a single result is applicable to all similar lattices with appropriate scaling, e.g., $\sigma_n^{(m)}(i)$ applies to all square lattices. Its explicit dependence on $\tau$ can be revealed by re-expressing \eqref{eq:sigma} as 
\begin{equation}
\sigma_n^{(m)}(\tau) = \sideset{}{'}\sum_{(p_1,p_2)\in\mathbb{Z}^2}\frac{(p_1+\tau^* p_2)^{(m-n)/2}}{(p_1+\tau p_2)^{(m+n)/2}},
\label{eq:sigmaeisen}
\end{equation}
where $\tau^*$ is the complex conjugate, exposing its similarity to conventional Eisenstein series. Similarly, the index of cylindrical harmonic functions \eqref{eq:Sdef} over the reciprocal lattice is given by $h = (h_1,h_2)$ and runs over $\overbar{\Omega}(\tau,a)$, containing the points $(K_h,\psi_h)$ in polar coordinates. 

%Exponentially convergent series are derived, valid for any $\tau$. Physically meaningful results even for conditionally convergent sums are available by regularization. For special lattice geometries, closed-form evaluations are obtained.

%The sums reduce to a simple polynomial in $u$, with coefficients $\sigma_n^{(m)}$.

\section{Evaluation of generalized Eisenstein sums}
\label{sec:eisenstein}
The sums \eqref{eq:sigma}, and the alternative form \eqref{eq:sigmaeisen}, are closely related to Eisenstein series and can be treated using associated techniques. By transforming to the Fourier domain, rapidly convergent series are derived convenient for numerical evaluation, given in Section \ref{sec:fourier}. Furthermore, using the theory of modular forms and elliptic functions, closed-form expressions for $m \geqslant n$ are available for several special geometries, including but not limited to the square, hexagonal, $2:1$ and $\sqrt{3}:1$ rectangular lattices, derived in Section \ref{sec:exact}. 

For $\sigma_2^{(m)}$, the sums are conditionally convergent, and its value depends critically on summation order. The non-physical contribution that arises differs for different summation techniques, known variously as a non-modular contribution or extraordinary contribution. However, the sum can be regularized by enforcing the geometric identity
\begin{equation}
\sigma_n^{(m)}(\tau) = \frac{|\tau|^{m-n}}{\tau^m}\sigma_n^{(m)}(-1/\tau),
\label{eq:invtau}
\end{equation}
corresponding to a rotation and scaling of the lattice. For $n>2$, \eqref{eq:invtau} holds automatically as the sum is absolutely convergent, which allows the identity to be derived by rearranging the summation indexes $p_1$ and $p_2$. The conditionally convergent case of $n=2$ is regularized in Section \ref{sec:nonmod} by comparing with a sum which does obey \eqref{eq:invtau}. In conjunction with the rapidly convergent Fourier series derived in Section \ref{sec:fourier}, this provides a robust yet simple means of obtaining physically meaningful results. 
\subsection{Conversion to Eisenstein series}
Eisenstein series are defined by the sum
\begin{equation}
G_n(\tau) = \sigma_n^{(n)}(\tau) = \sideset{}{'}\sum_{(p_1,p_2)\in\mathbb{Z}^2}\frac{1}{(p_1+\tau p_2)^n}.
\label{eq:eisenlattice}
\end{equation}
From the theory of modular forms, the double sum can be converted to a Fourier series,\autocite{stein2010complex,zagier1992introduction}
\begin{equation}
G_n(\tau) = 2\zeta(n) + \frac{2(2i\pi)^n}{(n-1)!} \sum_{r=1}^\infty \hat{\sigma}_{n-1}(r) e^{2i\pi r\tau},
\label{eq:eisenfour}
\end{equation}
where $\zeta(n)$ is the Riemann zeta function, and $\hat{\sigma}_k(r)$ is the sum of the $k$th power of the divisors of $r$,
\begin{equation}
\hat{\sigma}_k(r) = \sum_{d|r}d^k.
\end{equation}
Alternatively, the result can be expressed as a Lambert series, via\autocite{stein2010complex,mityushev2002longitudinal}
\begin{equation}
\sum_{r=1}^\infty \hat{\sigma}_k(r) q^r = \sum_{r=1}^\infty \frac{r^k q^r}{1-q^r}.
\end{equation}
However, we will use the Fourier series exclusively, as it accepts generalizations more readily.

\subsubsection{Fourier series for generalized Eisenstein series}
\label{sec:fourier}
We now generalize the result \eqref{eq:eisenfour} to $\sigma_n^{(m)}$, where $m$ and $n$ are even integers, obtaining a polynomial expression in $G_n(\tau)$ and its derivatives when $m>n$. We show the derivation explicitly for the case $m = n+2$, from which all other cases follow. Beginning with the definition, we first partition the sum, evaluating the contribution from the real axis separately, and use symmetry to obtain
\begin{equation}
\sigma_n^{(n+2)}(\tau) 
= \sideset{}{'}\sum_{(p_1,p_2)\in\mathbb{Z}^2}\frac{p_1+\tau^* p_2}{(p_1+\tau p_2)^{n+1}} 
= \sideset{}{'}\sum_{p_1\in\mathbb{Z}}\frac{1}{p_1^n} + 2\sum_{p_2=1}^\infty\sum_{p_1\in\mathbb{Z}}\frac{p_1+\tau^* p_2}{(p_1+\tau p_2)^{n+1}}.
\label{eq:splitsum}
\end{equation}
We perform the sum over $p_1$ first, which imposes the Eisenstein summation order of summation which is immaterial for absolutely convergent orders, but becomes critical for conditionally convergent orders. The first sum in \eqref{eq:splitsum} can be recognized as twice the Riemann zeta function, while a Fourier domain representation is sought for the sum over $p_1$ in the second sum via the Poisson summation formula,
\begin{equation}
\sum_{q\in\mathbb{Z}} f(q) = \sum_{r\in\mathbb{Z}} \hat{f}(r),
\label{eq:poissonm}
\end{equation}
where the Fourier transform is defined
\begin{equation}
\hat{f}(y) = \int_{-\infty}^\infty f(x)e^{-2i\pi xy} dx.
\end{equation}

Application of \eqref{eq:poissonm} to \eqref{eq:splitsum} requires consideration of the integral
\begin{equation}
\hat{f}(r) = \int_{-\infty}^\infty \frac{x+c^*}{(x+c)^k} e^{-2i\pi xr}dx.
\label{eq:fourint}
\end{equation}
From \eqref{eq:splitsum}, only the case where the pole of the integrand lies in the lower half plane is needed. For positive $r$, the integral is evaluated by selecting a semi-circular contour in the lower half plane and invoking Cauchy's residue theorem, yielding
\begin{equation}
\hat{f}(r)= \frac{(-2i\pi)^{k-1}}{(k-2)!} r^{k-2} e^{2i\pi rc} - 2i\imag(c)\frac{(-2i\pi)^k}{(k-1)!} r^{k-1} e^{2i\pi rc}.
\label{eq:fhatr}
\end{equation}
Non-positive values of $r$ allow a semi-circular contour in the upper half plane to be selected, which does not enclose any poles, so the result is zero. This result, known as a Lipschitz summation formula, has a general form that we later use.\autocite{maass1983lectures,pasles2001generalization}

The sum then takes the form
\begin{equation}
\sigma_n^{(n+2)}(\tau) 
= 2\zeta(n)+ 2\sum_{p_2=1}^\infty\sum_{r=1}^\infty \left[\frac{(2i\pi)^n}{(n-1)!} r^{n-1} e^{2i\pi r\tau p_2} + 2i\imag(\tau)\frac{(2i\pi)^{n+1}}{n!} r^n p_2 e^{2i\pi r\tau p_2}\right].
\end{equation}
The double sum can be expanded term by term, and converted to a single sum by gathering equal powers of $e^{2i\pi \tau}$ to give
\begin{equation}
\sigma_n^{(n+2)}(\tau) 
= 2\zeta(n)+ \frac{2(2i\pi)^n}{(n-1)!}\sum_{r=1}^\infty \hat{\sigma}_{n-1}(r) e^{2i\pi r\tau} + 2i\imag(\tau)\frac{2(2i\pi)^{n+1}}{n!} \sum_{r=1}^\infty r\hat{\sigma}_{n-1}(r) e^{2i\pi r\tau}.
\label{eq:fouriern2}
\end{equation}
The first sum, along with the contribution from the real axis, is the Fourier series of the fundamental Eisenstein series \eqref{eq:eisenfour}, while the second sum is its derivative, yielding the compact result
\begin{equation}
\sigma_n^{(n+2)}(\tau) = G_n(\tau) + \frac{2i\imag(\tau)}{n}G'_n(\tau),
\label{eq:sigman2}
\end{equation}
where the prime represents differentiation with respect to the argument. The case $m = n+4$ can be treated by similar means, to give
\begin{equation}
\sigma_n^{(n+4)}(\tau) = G_n(\tau) + \frac{4i\imag(\tau)}{n} G'_n(\tau) - \frac{4\imag(\tau)^2}{n(n+1)} G''_n(\tau).
\label{eq:sigman4}
\end{equation}

The general case can also be treated by similar means, or by using the generalized Lipschitz summation formula directly, which automates the task of evaluating the residues. Its most convenient form reads\autocite{maass1983lectures,pasles2001generalization}
\begin{equation}
\begin{split}
&\sum_{q=-\infty}^\infty \frac{1}{(q+c)^\alpha(q+c^*)^\beta} = \frac{(-i)^{\alpha-\beta}(2\pi)^{\alpha+\beta}}{\Gamma(\alpha)\Gamma(\beta)}\left[\Gamma(\alpha+\beta+1)(4\pi\imag(c))^{1-\alpha-\beta} \vphantom{\sum_r^\infty}\right.\\
&+ \sum_{r=1}^\infty \Gamma(\beta) r^{\alpha+\beta-1} U(\beta,\alpha+\beta,4\pi r\imag(c)) e^{2i\pi rc}
\left. + \sum_{r=1}^\infty \Gamma(\alpha) r^{\alpha+\beta-1} U(\alpha,\alpha+\beta,4\pi r\imag(c)) e^{-2i\pi rc^*}\right],
\end{split}
\end{equation}
where $U(\alpha,\beta,z)$ is the Tricomi confluent hypergeometric function. We proceed firstly for $m>n$, corresponding to negative $\beta$. Then only the first infinite sum survives, due to the pole $\Gamma(\beta)$ in the prefactor. Furthermore, expression in terms of a generalized Laguerre polynomial is possible,\autocite{abramowitz1964handbook}
\begin{equation}
U(-s,\beta,z) = (-1)^s s! L_s^{(\beta-1)}(z) = (-1)^s\sum_{t=0}^s\binom{s}{t}\frac{(s+\beta-1)!}{(t+\beta-1)!}(-z)^t.
\end{equation}
The procedure continues as before, requiring now the general resummation formula,
\begin{equation}
\sum_{p=1}^\infty\sum_{r=1}^\infty p^\alpha r^\beta q^{pr} = \sum_{r=1}^\infty r^\beta \hat{\sigma}_{\alpha-\beta}(r)q^r.
\end{equation}
Finally, we obtain the general result
\begin{equation}
\sigma_n^{(m)}(\tau) = \sum_{k=0}^{(m-n)/2} (2i\imag(\tau))^k \binom{(m-n)/2}{k} \frac{(n-1)!}{(n+k-1)!} \frac{\partial^k}{\partial\tau^k} G_n(\tau),
\label{eq:sigmanm}
\end{equation}
valid for all $m \geqslant n$, where both $m$ and $n$ are positive even integers. From \eqref{eq:eisenfour}, the Fourier expansion for the $k$th derivative of $G_n$ is
\begin{equation}
\frac{\partial^k}{\partial\tau^k} G_n(\tau) = \frac{2(2i\pi)^{n+k}}{(n-1)!} \sum_{r=1}^\infty r^k \hat{\sigma}_{n-1}(r) e^{2i\pi r\tau},
\label{eq:fourierderiv}
\end{equation}
with $G_n(\tau)$ itself having the additional constant term in \eqref{eq:eisenfour}.

These provide a numerically convenient means of evaluating $\sigma_n^{(m)}(\tau)$, for any lattice $\tau$, and especially the conditionally convergent $\sigma_2^{(m)}(\tau)$, which can subsequently be regularized by the methods of Section \ref{sec:nonmod}. The simple form of \eqref{eq:sigmanm}, as a polynomial in $G_n(\tau)$ and its derivatives, also serves as the basis for closed-form evaluations for several special lattices in Section \ref{sec:exact}. The Fourier series \eqref{eq:fourierderiv} converges as $\mathcal{O}(r^{n+k-1}q^r)$ as $r\rightarrow\infty$.\autocite{zagier1992introduction} Note that $|q| = |e^{2i\pi\tau}| <1$, and can be minimized by maximizing $\imag(\tau)$, via \eqref{eq:invtau}. Furthermore, series with convergence worse than $\mathcal{O}(r^5 q^r)$ can be avoided with the aid of recurrence relations \eqref{eq:eisenrecur} and \eqref{eq:eisenderiv}, though this is rarely necessary given they are rapidly converging series. As expected from an application of the Poisson summation formula, series which are the slowest to converge over the direct lattice are the fastest to converge in the Fourier domain. 

For completeness, we also provide the Fourier series of $\sigma_n^{(m)}(\tau)$ for $m < n$. Once again, the confluent hypergeometric function can be expressed as a generalized Laguerre polynomial, by use of the additional identity,\autocite{abramowitz1964handbook}
\begin{equation}
U(\alpha,\beta,z) = z^{1-\beta}U(\alpha-\beta+1,2-\beta,z).
\end{equation}
The procedure now yields
\begin{equation}
\begin{split}
\sigma_n^{(m)}(\tau) &= 2\zeta(n) + \binom{n-2}{\alpha-1} \frac{(-i)^m\pi}{2^{n-3}\imag(\tau)^{n-1}}\zeta(n-1)\\
&+\sum_{k=0}^{\alpha-1}\frac{(-i)^m\pi^{\alpha-k}}{2^{2k-m-1}\imag(\tau)^{\beta+k}}\binom{\beta+k-1}{k}\frac{1}{(\alpha-k-1)!}\sum_{r=1}^{\infty}\frac{1}{r^{\beta+k}}\hat{\sigma}_{n-1}(r)e^{2i\pi r\tau}\\
&+\sum_{k=0}^{\beta-1}\frac{(-i)^m\pi^{\beta-k}}{2^{2k+m-1}\imag(\tau)^{\alpha+k}}\binom{\alpha+k-1}{k}\frac{1}{(\beta-k-1)!}\sum_{r=1}^{\infty}\frac{1}{r^{\alpha+k}}\hat{\sigma}_{n-1}(r)e^{-2i\pi r\tau^*},
\end{split}
\label{eq:hyper}
\end{equation}
where explicitly the variables $\alpha = (n+m)/2$ and $\beta = (n-m)/2$. 

Although \eqref{eq:hyper} is still expressed as a polynomial of functions related to $G_n(\tau)$, it is of less practical utility than \eqref{eq:sigmanm}, for two reasons. Firstly, the sum over $r$ is related to the antiderivative of $G_n(\tau)$ rather than its derivative, and does not lead to closed-form evaluations by the techniques of Section \ref{sec:exact}. Instead, alternative number theoretic techniques have been employed by other authors to yield exact results, whose results we quote in Section \ref{sec:S0}. Secondly, the case $m < n$ can be summed from its definition \eqref{eq:sigmaeisen} to sufficient accuracy using modern computing resources with little difficulty, as all the sums are absolutely convergent over the direct lattice. The sole exception is $\sigma_2^{(0)}(\tau)$, which is divergent. This is reflected in \eqref{eq:hyper}, where the $\zeta(n-1)$ term has a simple pole at $n=2$, and its residue is in accordance with Kronecker's first limit formula.\autocite{siegel1961advanced,weil2012elliptic} In Section \ref{sec:S0}, a particular regularization method for this divergent sum based on Kronecker's second limit formula is employed.

\subsubsection{Regularization of conditionally convergent generalized Eisenstein series}
\label{sec:nonmod}
As a double sum over the lattice, $\sigma_2^{(m)}(\tau)$ is conditionally convergent, so \eqref{eq:sigma} is ill-defined unless a particular order of summation is enforced. We follow the convention in the modular forms literature, corresponding to the Eisenstein summation method, of first summing over index $p_1$ then over $p_2$. For $G_2(\tau)$, this produces the value given by its absolutely convergent Fourier series \eqref{eq:eisenfour}, \autocite{stein2010complex,weil2012elliptic} while for a general order $\sigma_2^{(m)}(\tau)$ we obtained \eqref{eq:sigmanm}. However, another problem persists: the transformation $\tau \rightarrow -1/\tau$, corresponding to a rotation and a scaling of the lattice, fails to satisfy the identity \eqref{eq:invtau}. On geometrical grounds, this transformation should be equivalent to scaling by a complex factor, and physically meaningful results for the infinite lattice are only possible if \eqref{eq:invtau} is obeyed.

Such a series can be regularized, defined by analytical continuation from \eqref{eq:sigmaeisen} and taking the limit,
\begin{equation}
\tilde{\sigma}_2^{(m)}(\tau) = \lim_{s\rightarrow0^+}\sideset{}{'}\sum_{(p_1,p_2)\in\mathbb{Z}^2}\frac{(p_1+\tau^* p_2)^{(m-2)/2}}{(p_1+\tau p_2)^{(m+2)/2}|p_1+\tau p_2|^s}.
\label{eq:sigmaac}
\end{equation}
This is a two-dimensional analogue of regularization by the analytical continuation of the Riemann zeta function.\autocite{hawking1977zeta,elizalde1994zeta} The series now transforms in a physical way,
\begin{equation}
\tilde{\sigma}_2^{(m)}(\tau) = \frac{|\tau|^{m-2}}{\tau^m}\tilde{\sigma}_2^{(m)}(-1/\tau),
\label{eq:invtau2}
\end{equation}
and has physically meaningful values. Using Hecke's trick,\autocite{zagier1992introduction} $\tilde{\sigma}_2^{(m)}(\tau)$ and $\sigma_2^{(m)}(\tau)$ are shown to differ only by a constant, so \eqref{eq:sigmaac} never needs to be explicitly evaluated since the non-physical contribution can easily be identified and subtracted from \eqref{eq:sigmanm}.

%We demonstrate this in Section \ref{sec:gamma}, by successfully expressing absolutely convergent sums over the reciprocal lattice in terms of the regularized $\tilde{\sigma}_2^{(m)}(\tau)$. 

The procedure proceeds by retracing the initial steps of Section \ref{sec:fourier}, beginning by partitioning the sum \eqref{eq:sigmaac} as in \eqref{eq:splitsum}, and applying the Poisson summation formula \eqref{eq:poissonm}. As in the previous section, we show the derivation explicitly only for $m=4$, and state the general result. We thus evaluate an integral similar to \eqref{eq:fourint},
\begin{equation}
\hat{f}_{p_2}(r) = \int_{-\infty}^\infty \frac{x+\tau^* p_2}{(x+\tau p_2)^3|x+\tau p_2|^s} e^{-2i\pi xr} dx.
\label{eq:fourintac}
\end{equation}
For $r>0$, the result of the integral becomes identical to \eqref{eq:fhatr}, in the limit $s \rightarrow 0^+$. But, for $r=0$, \eqref{eq:fourintac} does not approach $0$, which is the corresponding value in \eqref{eq:fhatr}. To evaluate this additional contribution, we transform the integral with $x \rightarrow p_2(x\imag(\tau) - \real(\tau))$ to give
\begin{equation}
\hat{f}_{p_2}(0) = \frac{1}{(p_2\imag(\tau))^{1+s}}\int_{-\infty}^\infty \frac{1}{(x+i)^{3+\frac{s}{2}}(x-i)^{-1+\frac{s}{2}}} dx.
\end{equation}
Two successive partial integrations give
\begin{equation}
\hat{f}_{p_2}(0) = \frac{1}{(p_2\imag(\tau))^{1+s}} \frac{s(s-2)}{(s+2)(s+4)}\int_{-\infty}^\infty \frac{1}{(x^2+1)^{1+\frac{s}{2}}} dx.
\end{equation}

Next, the sum over $p_2$ is evaluated,
\begin{equation}
2\sum_{p_2=1}^\infty \hat{f}_{p_2}(0) = \frac{s\zeta(1+s)}{\imag(\tau)^{1+s}} \frac{2(s-2)}{(s+2)(s+4)}\int_{-\infty}^\infty \frac{1}{(x^2+1)^{1+\frac{s}{2}}} dx.
\end{equation}
In the limit $s \rightarrow 0^+$, $s\zeta(1+s)$ approaches $1$, corresponding to the residue of the Riemann zeta function at its pole, while the integral evaluates to $\pi$, yielding the result
\begin{equation}
\tilde{\sigma}_2^{(4)}(\tau) = \sigma_2^{(4)}(\tau) - \frac{\pi}{2\imag(\tau)}.
\end{equation}
After some algebraic manipulations, the general result for $m$ greater than $2$ by an even integer is
\begin{equation}
\tilde{\sigma}_2^{(m)}(\tau) = \sigma_2^{(m)}(\tau) - \frac{2\pi}{m\imag(\tau)}.
\label{eq:nonmod}
\end{equation}
This identity obtains the regularized $\tilde{\sigma}_2^{(m)}(\tau)$ from $\sigma_2^{(m)}(\tau)$, eliminating the non-physical extraordinary contribution from any summation procedure conforming to the Eisenstein summation order, which includes all results obtained in this paper. Finally, \eqref{eq:nonmod} can be combined with \eqref{eq:invtau2} to yield a transformation identity for non-regularized sums,
\begin{equation}
\sigma_2^{(m)}(\tau) = \frac{|\tau|^{m-2}}{\tau^m}\sigma_2^{(m)}(-1/\tau) + \frac{2\pi}{m\imag(\tau)}\left[1 - \frac{|\tau|^m}{\tau^m}\right].
\label{eq:invtau2m}
\end{equation}

\subsection{Exact values of generalized Eisenstein series}
\label{sec:exact}
The procedure for obtaining closed-form solutions of $\sigma_n^{(m)}(\tau)$ is comprised of three steps. The first step, already performed in Section \ref{sec:fourier}, was to express the sums as linear combinations of Eisenstein series and its derivatives \eqref{eq:sigmanm}. Secondly, using derivative and recurrence relations for the Eisenstein series, all results can then be expressed in terms of only three fundamental Eisenstein series. Finally, results can be converted to combinations of Dedekind $\eta$-functions and Weber $\eta$-quotients, for which explicit values are available for many lattices.

All integer orders of $G_n(\tau)$ and all its derivatives can be reduced to sums and products of $G_2(\tau)$, $G_4(\tau)$, and $G_6(\tau)$, using basic results from the theory of modular forms. For example, the recurrence relations for the lowest orders are
\begin{align}
7G_8 &= 3G_4^2, & 11G_{10} &= 5G_4G_6, & 143G_{12} &= 18G_4^3 + 25G_6^2,
\end{align}
suppressing the $\tau$ dependence for brevity. A general recursion relation is available for all $G_{2k}(\tau)$,\autocite{goursat1916functions,mityushev2002longitudinal}
\begin{equation}
G_{2k} = \frac{3}{(2k+1)(2k - 1)(k - 3)} \sum_{s=2}^{k-2} (2s-1)(2k - 2s - 1) G_{2s} G_{2(k-s)}.
\label{eq:eisenrecur}
\end{equation}

Meanwhile, $G_2(\tau)$, $G_4(\tau)$, and $G_6(\tau)$ form a closed ring under differentiation,
\begin{align}
2i\pi G'_2 &= 5G_4 - G_2^2, & i\pi G'_4 &= 7G_6 - 2G_2G_4, & 7i\pi G'_6 = 30G_4^2 - 21G_2G_6,
\label{eq:eisenderiv}
\end{align}
obtained by Ramanujan,\autocite{ramanujan1916certain} and thus all higher order derivatives can be obtained by successive application. These relations are correct when $G_2(\tau)$ is defined by its absolutely convergent Fourier series \eqref{eq:eisenfour}, rather than its conditionally convergent sum over the lattice \eqref{eq:eisenlattice}. 

\subsubsection{Closed form evaluation of $G_4(\tau)$ and $G_6(\tau)$}
Having decomposed $\sigma_n^{(m)}(\tau)$ entirely in terms of $G_2(\tau)$, $G_4(\tau)$, and $G_6(\tau)$, it remains only to evaluate these conventional Eisenstein series for the lattices of interest, $\tau$. While several procedures are now possible, we opt to further decompose these series in terms of Dedekind $\eta$-functions, for which many special values are known and the Chowla--Selberg formula and its generalizations can be used to generate many more.\autocite{chowla1949epsteins,zucker1977evaluation,vanderpoorten1999values,hart2015class}

From the theory of elliptic functions, it is known that $G_4(\tau)$ and $G_6(\tau)$ are related to the invariants of the Weierstrass $\wp$-function, which can be expressed as Jacobi $\vartheta$-functions,\autocite{abramowitz1964handbook}
\begin{equation}
\begin{gathered}
G_4(\tau) = \frac{\pi^4}{90}[\vartheta_2(0;\tau)^8 + \vartheta_3(0;\tau)^8 + \vartheta_4(0;\tau)^8],\\
33075G_6(\tau)^2 = 13500G_4(\tau)^3 - \pi^{12}[\vartheta_2(0;\tau)\vartheta_3(0;\tau)\vartheta_4(0;\tau)]^8.
\end{gathered}
\end{equation} 
These are related to the Dedekind $\eta$-functions via the identities
\begin{align}
\vartheta_2(0;\tau) &= \frac{2\eta(2\tau)^2}{\eta(\tau)}, &
\vartheta_3(0;\tau) &= \frac{\eta\left(\frac{1}{2}(\tau+1)\right)^2}{\eta(\tau+1)}, &
\vartheta_4(0;\tau) &= \frac{\eta(\tfrac{1}{2}\tau)^2}{\eta(\tau)}.
\end{align} 
To limit the number of Dedekind $\eta$-function values necessary, we also introduce Weber's $\eta$-quotients,\autocite{weber1908lehrbuch}
\begin{align}
\mathfrak{f}(\tau) &= e^{-\frac{i\pi}{24}}\frac{\eta\left(\frac{1}{2}(\tau+1)\right)}{\eta(\tau)}, &
\mathfrak{f}_1(\tau) &= \frac{\eta(\frac{1}{2}\tau)}{\eta(\tau)}, &
\mathfrak{f}_2(\tau) &= \sqrt{2}\frac{\eta(2\tau)}{\eta(\tau)},
\label{eq:weber}
\end{align}
which are algebraic numbers for all $\tau = \sqrt{-c}$, where $c$ is a positive integer. Furthermore, all the $\eta$-quotients are interrelated by the simple formulas
\begin{align}
\mathfrak{f}(\tau)\mathfrak{f}_1(\tau)\mathfrak{f}_2(\tau) &= \sqrt{2}, &
\mathfrak{f}(\tau)^8 &= \mathfrak{f}_1(\tau)^8 + \mathfrak{f}_2(\tau)^8.
\label{eq:weberid}
\end{align}

The $\vartheta$-functions are now expressed as
\begin{align}
\vartheta_2(0;\tau) &= \eta(\tau)\mathfrak{f}_2(\tau)^2, &
\vartheta_3(0;\tau) &= \eta(\tau)\mathfrak{f}(\tau)^2, &
\vartheta_4(0;\tau) &= \eta(\tau)\mathfrak{f}_1(\tau)^2.
\label{eq:thetaweber}
\end{align} 
These relations incorporate the functional equations for $\eta(\tau)$,
\begin{align}
\eta(\tau+1) &= e^{\frac{i\pi}{12}}\eta(\tau), & \eta(-1/\tau) &= \sqrt{-i\tau}\eta(\tau).
\label{eq:etamod}
\end{align}
Simpler expressions for $G_4(\tau)$ and $G_6(\tau)$ are thus obtained,
\begin{align}
G_4(\tau) &= \frac{\pi^4}{45}[\mathfrak{f}(\tau)^{16} - 16\mathfrak{f}(\tau)^{-8}]\eta(\tau)^8, & 33075G_6(\tau)^2 &= 13500G_4(\tau)^3 - 256\pi^{12}\eta(\tau)^{24},
\label{eq:etaweber}
\end{align}
where the latter is also a consequence of the vector space of modular forms.\autocite{stein2010complex,zagier1992introduction}

Up to now, the procedure is general to any lattice. We now proceed for the square ($\tau = i$) and hexagonal ($\tau = (1 + \sqrt{3}i)/2$) lattices, duplicating the well-known values of $G_4(i)$ and $G_6(e^{i\pi/3})$.\autocite{waldschmidt2008elliptic} We also treat the rectangular versions of these lattices ($\tau = 2i$, $\sqrt{3}i$), where $\tau = 2i$ is constructed by doubling one of the two lattice vectors. We remark that the equivalence of certain lattices, such as $\tau = 1+\sqrt{3}i$ and $\tau = \sqrt{3}i$, is a consequence of the periodicity in $\tau$,
\begin{equation}
\sigma_n^{(m)}(\tau+1) = \sigma_n^{(m)}(\tau),
\label{eq:tauperiod}
\end{equation}
for all $\tau$. This corresponds to translating each row of the lattice by one lattice vector horizontally, yielding an identical lattice.

Only two fundamental values of $\eta(\tau)$ are needed for all lattices considered,
\begin{align}
\eta(i) &= \frac{\Gamma(\frac{1}{4})}{2\pi^{3/4}}, & \eta(\sqrt{3}i) &= \frac{3^{1/8}\Gamma(\frac{1}{3})^{3/2}}{2^{4/3}\pi},
\label{eq:etasquare}
\end{align}
along with the values of $\eta$-quotients given by Weber,
\begin{align}
\mathfrak{f}(i) &= 2^{1/4}, & \mathfrak{f}(\sqrt{3}i) &= 2^{1/3}. 
\end{align}
Then the other relevant values of $\eta(\tau)$ can be deduced using a combination of \eqref{eq:weber}, \eqref{eq:weberid}, and \eqref{eq:etamod}, 
\begin{align}
\eta(2i) &= \frac{\Gamma(\frac{1}{4})}{2^{11/8}\pi^{3/4}}, & \eta(e^{i\pi/3}) &= e^{i\pi/24}\frac{3^{1/8}\Gamma(\frac{1}{3})^{3/2}}{2\pi},
\label{eq:etarecthex}
\end{align}
as well as
\begin{align}
\mathfrak{f}(2i) &= (4+3\sqrt{2})^{1/8}, & \mathfrak{f}(e^{i\pi/3}) &= 2^{1/6}.
\end{align}
By \eqref{eq:etaweber}, this yields all values of $G_4(\tau)$ and $G_6(\tau)$ on Table \ref{tab:sigma}. 

\begin{table}[t]
\centering
\begin{tabular}{>{$}c<{$} | >{$}c<{$} >{$}c<{$} >{$}c<{$} >{$}c<{$} >{$}c<{$} >{$}c<{$}}
\toprule[1pt]
& G_2 \equiv \sigma_2^{(2)} & G_4 \equiv \sigma_4^{(4)} & \sigma_2^{(4)} & G_6 \equiv \sigma_6^{(6)} & \sigma_4^{(6)} & \sigma_2^{(6)}\\
\midrule
\tau = i & 0 + \pi & \frac{\Gamma(\frac{1}{4})^8}{960\pi^2} & \frac{\Gamma(\frac{1}{4})^8}{384\pi^3}+\frac{\pi}{2} & 0 & 0 & 0 + \frac{\pi}{3} \\
\tau = 2i & \frac{\Gamma(\frac{1}{4})^4}{32\pi} + \frac{\pi}{2} & \frac{11\Gamma(\frac{1}{4})^8}{15360\pi^2} & \frac{\Gamma(\frac{1}{4})^8}{384\pi^3}+\frac{\pi}{4} & \frac{\Gamma(\frac{1}{4})^{12}}{81920\pi^3} & \frac{\Gamma(\frac{1}{4})^{12}}{24576\pi^4} & \frac{\Gamma(\frac{1}{4})^{12}}{6144\pi^5} + \frac{\pi}{6}\\
\tau = e^{i\pi/3} & 0 + \frac{2\pi}{\sqrt{3}} & 0 & 0 + \frac{\pi}{\sqrt{3}} & \frac{\Gamma(\frac{1}{3})^{18}}{8960\pi^6} & \frac{\sqrt{3}\Gamma(\frac{1}{3})^{18}}{5120\pi^7} & \frac{\Gamma(\frac{1}{3})^{18}}{1024\pi^8} + \frac{2\pi}{3\sqrt{3}}\\
\tau = \sqrt{3}i & \frac{\Gamma(\frac{1}{3})^6}{16 \cdot 2^{2/3} \pi^2} + \frac{\pi}{\sqrt{3}} & \frac{\Gamma(\frac{1}{3})^{12}}{512 \cdot 2^{1/3} \pi^4} & \frac{\sqrt{3}\Gamma(\frac{1}{3})^{12}}{256 \cdot 2^{1/3} \pi^5}+\frac{\pi}{2\sqrt{3}} & \frac{11\Gamma(\frac{1}{3})^{18}}{286720\pi^6} & \frac{3\sqrt{3}\Gamma(\frac{1}{3})^{18}}{40960\pi^7} & \frac{\Gamma(\frac{1}{3})^{18}}{2048\pi^8} + \frac{\pi}{3\sqrt{3}}\\
\bottomrule[1.2pt]
\end{tabular}
\caption{Exact values of some $\sigma_n^{(m)}$ for square, hexagonal, and rectangular lattices. Where two terms appear, the second term is the non-physical extraordinary contribution. Regularization of these results by \eqref{eq:nonmod} amounts to simply neglecting this second term, as demonstrated when applied to \eqref{eq:Sm}. For rectangular lattices, values for the opposite aspect ratio can be derived using \eqref{eq:invtau}, or using \eqref{eq:invtau2m} if an extraordinary contribution is present.}
\label{tab:sigma}
\end{table}

\subsubsection{Closed form evaluation of $G_2(\tau)$}
$G_2(\tau)$ can be expressed as the logarithmic derivative of $\eta(\tau)$,\autocite{stein2010complex,zagier1992introduction}
\begin{equation}
G_2(\tau) = -4i\pi\frac{\eta'(\tau)}{\eta(\tau)},
\end{equation}
where, again, $G_2(\tau)$ is defined via its Fourier series \eqref{eq:eisenfour}. However, neither $G_2(\tau)$ nor $\eta'(\tau)$ are modular forms, and fewer theoretical tools are available to evaluate $\eta'(\tau)$. Instead, we resort to a case by case treatment of $G_2(\tau)$ for the lattices of interest. For the square and hexagonal lattices, $G_2(\tau)$ would be zero by symmetry, but these values are non-zero due to the non-modular contribution. Thus, they can be derived from \eqref{eq:nonmod}, and their values are displayed on Table \ref{tab:sigma}. These agree with values previously reported in the literature.\autocite{perrins1979transport,mityushev2002longitudinal}

We focus attention on the last remaining values of interest, $G_2(2i)$ and $G_2(\sqrt{3}i)$, exploiting the fact that $G_2(\tau) - sG_2(s\tau)$ is a modular form on a subset of the integers. In particular,
\begin{equation}
\vartheta_3(0;\tau)^4 = -\frac{1}{\pi^2}[G_2(\tau/2) - 4G_2(2\tau)],
\label{eq:foursquare}
\end{equation}
an identity used to prove the four squares theorem.\autocite{stein2010complex,zagier1992introduction} Using \eqref{eq:invtau2m}, it immediately follows that
\begin{equation}
G_2(2i) = \frac{\pi^2}{8}\vartheta_3^4(0,i) + \frac{\pi}{2} = \frac{\Gamma(\frac{1}{4})^4}{32\pi} + \frac{\pi}{2},
\end{equation}
which was evaluated using \eqref{eq:thetaweber}. Intriguingly, \eqref{eq:foursquare} can also be employed to evaluate $G_2(\sqrt{3}i)$, utilizing both \eqref{eq:invtau2m} and \eqref{eq:tauperiod} to obtain
\begin{equation}
G_2(2e^{i\pi/3}) = \frac{\pi^2}{2(3+\sqrt{3}i)}\vartheta_3^4(0,e^{i\pi/3}) + \frac{\pi}{\sqrt{3}} = \frac{\Gamma(\frac{1}{3})^6}{16 \cdot 2^{2/3} \pi^2} + \frac{\pi}{\sqrt{3}}.
\end{equation}

These exact values complete the procedure required to evaluate the exact values on Table \ref{tab:sigma}, which compare with previously obtained numerical values.\autocite{perrins1979transport,helsing1994bounds,berman1994renormalization,movchan1997greens,mityushev2002longitudinal} Along with the values for $G_4(\tau)$ and $G_6(\tau)$ on Table \ref{tab:sigma}, values of $G_{2k}(\tau)$ and all its derivatives can be obtained using the recurrence relation \eqref{eq:eisenrecur} and Ramanujan's derivatives identities \eqref{eq:eisenderiv}, thus evaluating \eqref{eq:sigmanm} in closed form. Results for all lattices where Dedekind $\eta$-function evaluations are available can similarly be deduced for all even integers $m$ and $n$, where $m \geqslant n$.

\section{Evaluation of cylindrical harmonic sums}
\label{sec:gamma}
The results of Section \ref{sec:eisenstein} are now applied to obtain \eqref{eq:Sdef} over the lattice \eqref{eq:rlattice},
\begin{equation}
S_{l,m,n}(u;\tau,a) = \sideset{}{'}\sum_h \frac{J_l(K_hu)}{K_h^n}e^{im\psi_h},
\end{equation}
corresponding to all the $\Gamma$ points in each Brillouin zone, excluding the origin. The other high symmetry points will be considered later in Section \ref{sec:displaced}. When $l-n$ is even, $S_{l,m,n}(u;\tau,a)$ can be expressed as a terminating polynomial in $u$, with $\sigma_n^{(m)}$ as coefficients.\autocite{nicorovici1996analytical} When $l-n$ is odd, $S_{l,m,n}(u;\tau,a)$ must instead by expressed as an infinite series. Fortunately, even $l-n$ is the most commonly encountered case, and is the only case we consider here. Furthermore, only even $m$ are considered, as the sum is identically zero for any lattice centered at $\Gamma$ when $m$ is odd.

A convenient feature of $S_{l,m,n}(u;\tau,a)$ is that all orders $(l,n)$ of a given $m$ are linked by recurrence relations. Thus, the strategy for evaluating all $(l,m,n)$ begins by finding an expression for a particular $(l,n)$ for each $m$, then successively applying the recurrence relations.

\subsection{Recurrence relations}
\label{sec:recur}
Bessel functions obey a set of recurrence relations which link differing orders, yielding recurrence relations derived by Nicorovici et al.\ for square lattices.\autocite{nicorovici1996analytical} Due to their importance, we restate these results in the present notation,
\begin{gather}
u^l S_{l-1,m,n-1}(u;\tau,a) = \partial_u [u^l S_{l,m,n}(u;\tau,a)],\label{eq:recur1}\\
u^{-l} S_{l+1,m,n-1}(u;\tau,a) = -\partial_u [u^{-l} S_{l,m,n}(u;\tau,a)],\\
u^{1+l} S_{l+1,m,n+1}(u;\tau,a) = \int_0^u v^{1+l} S_{l,m,n}(v;\tau,a) dv, \label{eq:recur3}\\
u^{1-l} S_{l-1,m,n+1}(u;\tau,a) = \frac{i^m}{2^{l-1} (l-1)!} \left(\frac{A}{2\pi a}\right)^{n-l+2} \sigma_{n-l+2}^{(m)}(\tau) - \int_0^u v^{1-l} S_{l,m,n}(v;\tau,a) dv.
\label{eq:recur4}
\end{gather}
Only \eqref{eq:recur4} depends on the lattice, and thus required generalization to general lattices. This begins with
\begin{equation}
\int_0^a x^{-l+1} J_l(bx) dx = \frac{b^{l-2}}{2^{l-1}(l-1)!} - a^{-l+1} \frac{J_{l-1}(ba)}{b},
\end{equation}
where $a$ and $b$ are arbitrary constants. Manipulating the identity gives
\begin{equation}
\int_0^u v^{1-l} \sideset{}{'}\sum_h \frac{J_l(K_h v)}{K_h^n} e^{im\psi_h} dv = \frac{1}{2^{l-1}(l-1)!} \sideset{}{'}\sum_h \frac{e^{im\psi_h}}{K_h^{n-l+2}} - u^{1-l} \sideset{}{'}\sum_h \frac{J_{l-1}(K_h u)}{K_h^{n+1}} e^{im\psi_h}.
\end{equation}
The sums are related to $S_{l,m,n}(u;\tau,a)$ and $\sigma_n^{(m)}(\tau)$, but summed over the reciprocal lattice. The geometrical relationship between direct \eqref{eq:lattice} and reciprocal \eqref{eq:rlattice} lattices then yields \eqref{eq:recur4}.

\subsection{Angle independent order ($m=0$)}
\label{sec:S0}
The case $m=0$ requires special treatment as it would otherwise feature divergent $\sigma_2^{(0)}(\tau)$ sums, and its regularization is considered in this section. We follow Nicorovici et al.,\autocite{nicorovici1996analytical} generalizing expressions for $S_{l,0,n}(u;\tau,a)$ to an arbitrary lattice $\overbar{\Omega}(\tau,a)$. The procedure begins with $S_{0,0,2}(u;\tau,a)$, which can be obtained from Kronecker's second limit formula,\autocite{siegel1961advanced,weil2012elliptic} which was rederived by Glasser and Stremler.\autocite{glasser1974evaluation,stremler2004evaluation} It reads
\begin{equation}
\sideset{}{'}\sum_h \frac{e^{i\bv{K}_h\cdot \bv{u}}}{K_h^2} = \frac{u^2}{4}(1-\cos2\varphi) - \frac{A}{6\pi}\log2 - \frac{A}{2\pi}\log\left|\frac{\vartheta_1(z/a;\tau)}{[\vartheta'_1(0;\tau)]^{1/3}}\right|,
\label{eq:stremler}
\end{equation}
where $\bv{K}_h = (K_h,\psi_h)$. Furthermore, $\bv{u} = x\bv{\hat{x}} + y\bv{\hat{y}}$ is an arbitrary vector in the unit cell, $z = x + iy$ is its complex representation, and $\varphi$ is its complex angle. The Jacobi $\vartheta_1$-function and its derivative is defined in concordance with Ref.\ \parencite{sansone1960lectures}. The desired identity follows by considering only the angle independent terms of \eqref{eq:stremler} to extract $S_{0,0,2}(u;\tau,a)$. The Jacobi--Anger identity allows the summand to be cast in terms of Bessel functions,
\begin{equation}
e^{i\bv{K}_h \cdot \bv{u}} = \sum_{l = -\infty}^\infty i^l J_l(K_h u) e^{il(\varphi-\psi_h)}.
\end{equation}
Meanwhile, the logarithm of the $\vartheta_1$-function is expanded as an asymptotic series, for which the real part reads
\begin{equation}
\log|\vartheta_1(z/a;\tau)| = \log u + \log\left(\frac{\pi|\vartheta'_1(0;\tau)|}{a}\right) + \frac{1}{6}\left(\frac{\pi}{a}\right)^2\left|\frac{\vartheta'''_1(0;\tau)}{\vartheta'_1(0;\tau)}\right|u^2\cos(2\varphi+\theta_\tau) + \cdots,
\end{equation}
where $\theta_\tau$ is the argument of $\vartheta'''_1(0;\tau)/\vartheta'_1(0;\tau)$, which depends only on $\tau$. Higher order terms of the asymptotic series all vary as $z^s$, for which the real part varies as $u^s\cos(s\varphi)$.

Dropping all terms that depend on $\varphi$ from \eqref{eq:stremler}, we obtain
\begin{equation}
S_{0,0,2}(u;\tau,a) = -\frac{A}{2\pi}\left[\log(u)+\log\left(\frac{2\pi\left|\eta(\tau)^2\right|}{a}\right)\right]+\frac{u^2}{4},
\label{eq:S002}
\end{equation}
where the Dedekind $\eta$-function is introduced via
\begin{equation}
\vartheta'_1(0;\tau) = 2\eta(\tau)^3.
\label{eq:thetaeta}
\end{equation}
Exact results are possible using special values of $\eta(\tau)$, given in \eqref{eq:etasquare} and \eqref{eq:etarecthex}. Otherwise, it may be numerically evaluated by a variety of efficient methods, such as by \eqref{eq:thetaeta} or by its Fourier series expansion,
\begin{equation}
\log(\eta(\tau)) = \frac{i\pi\tau}{12} - \sum_{r=1}^\infty \frac{1}{r}\hat{\sigma}_1(r) e^{2i\pi r\tau}.
\label{eq:logeta}
\end{equation}
Comparing \eqref{eq:logeta} with \eqref{eq:eisenfour} we observe that the $\log|\eta(\tau)|$ term plays the role of the divergent $\sigma_2^{(0)}(\tau)$ sum.

Expressions for all other $(l,n)$ follow from recurrence relations \eqref{eq:recur1}--\eqref{eq:recur4}, and do not require special treatment, since all additional terms feature only absolutely convergent sums. For orders $l < n$, expressions have the common form 
\begin{equation}
\begin{split}
S_{l,0,n}(u;\tau,a) = &\sum_{k=2}^{(n-l)/2} \left(\frac{A}{2\pi}\right)^{2k} B_{2k}(l,n) u^{n-2k} \frac{\sigma_{2k}^{(0)}(\tau)}{a^{2k}}\\
&- \frac{A}{2\pi} B_2(l,n) u^{n-2} \left[\log(u) + C_l(l,n) + \log\left(\frac{2\pi\left|\eta(\tau)^2\right|}{a}\right)\right] - \frac{1}{8} C_0(l,n) u^n.
\label{eq:S0upper}
\end{split}
\end{equation}
If $(n-l)/2 < 2$, then the first sum does not contribute. For orders $l \geqslant n$, $S_{l,m,n}(u;\tau,a)$ has a special form due to the logarithmic term originating from \eqref{eq:S002}. The orders $l=n$ have the common form
\begin{equation}
S_{l,0,n}(u;\tau,a) = \frac{A}{2\pi} B_l(l,n) u^{n-2} - \frac{1}{8} C_0(l,n) u^n.
\label{eq:S0diag}
\end{equation}
Finally, for orders $l > n$ the sums are given by
\begin{equation}
S_{l,0,n}(u;\tau,a) = \frac{A}{2\pi} B_l(l,n) u^{n-2}.
\label{eq:S0lower}
\end{equation}

The coefficients $B_{2k}(l,n)$ and $C_{2k}(l,n)$ are derived from successive application of the recurrence relations, with values
\begin{align}
B_{2k}(l,n) &= \frac{(-1)^k i^{n-l} 2^{2k-n}}{(\frac{n+l}{2} -k)! (\frac{n-l}{2}-k)!},\label{eq:Bcoeff}\\
C_{2k}(l,n) &= \frac{i^{n-l} 2^{2-n} k! (k+2)!}{(\frac{n+l}{2}+k)! (\frac{n-l}{2}+k)!},\label{eq:Ccoeff}
\end{align}
The coefficient $B_{2k}$ is undefined if $n-l < 2k$ and $C_{2k}$ is undefined if $l-n < 2k$. However, these undefined orders never contribute to \eqref{eq:S0upper} or to any subsequent expression which uses \eqref{eq:Bcoeff} and \eqref{eq:Ccoeff}. Alternatively, if the factorials in \eqref{eq:Bcoeff} and \eqref{eq:Ccoeff} are rewritten using Gamma functions, these expressions will automatically produce a zero result whenever appropriate. The remaining coefficients are associated with integration and differentiation of the logarithmic term, with
\begin{equation}
B_l(l,n) = \frac{(\frac{l-n}{2})!}{2^{n-1}(\frac{l+n}{2}-1)!},
\end{equation}
and
\begin{equation}
C_l(l,n) = -\frac{1}{2} \left[H\left(\frac{n+l}{2} - 1\right) + H\left(\frac{n-l}{2} -1 \right)\right].
\label{eq:logcoeff}
\end{equation}
where $H(s)$ is the $s$th harmonic number, given by
\begin{equation}
H(s) = \sum_{t=1}^s \frac{1}{t}.
\end{equation}

An example of the simple form of \eqref{eq:S0diag} for $l = n$ is
\begin{equation}
S_{2,0,2}(u,\tau,a) = \frac{A}{4\pi} - \frac{u^2}{8}.
\end{equation}
We also give an example for $l < n$,
\begin{equation}
S_{1,0,5}(u,\tau,a) = \left(\frac{A}{2\pi}\right)^4 \frac{u}{2} \frac{\sigma_4^{(0)}(\tau)}{a^4} + \frac{A}{2\pi}\frac{u^3}{16}\left[\log(u) - \frac{5}{4} + \log\left(\frac{2\pi\left|\eta(\tau)^2\right|}{a}\right)\right] - \frac{u^5}{384}.
\end{equation}

The generalized Eisenstein sums $\sigma_{2s}^{(0)}(\tau)$ are Epstein zeta functions and have been evaluated exactly in terms of Dirichlet $L$-series extending back to the work of Lorenz and later Hecke,\autocite{fletcher1946index,glasser1973evaluation}
\begin{align}
\sigma_{2s}^{(0)}(i) &= 4\zeta(s)\beta(s), & \sigma_{2s}^{(0)}(2i) &= 2(1-2^{-s}+2^{1-2s})\zeta(s)\beta(s),\\
\sigma_{2s}^{(0)}(e^{i\pi/3}) &= 6\zeta(s)g(s), & \sigma_{2s}^{(0)}(\sqrt{3}i) &= 2(1+2^{1-2s})\zeta(s)g(s).
\end{align}
Here, $\zeta(s)$ is the Riemann zeta function, $\beta(s)$ is the Dirichlet beta function, and $g(s)$ is one of the next most simple Dirichlet $L$-series,
\begin{equation}
g(s) = 1-2^{-s}+4^{-s}-5^{-s}+7^{-s}\ldots
\end{equation}
Further results for a wide variety of lattices have been tabulated.\autocite{borwein2013lattice} While many efficient series and integral representations are available for these Dirichlet $L$-series, in practice exact values or decimal approximations are well known for small integer $s$. For example, 
\begin{align}
\lambda &\equiv \beta(2) = 0.915965594177219\ldots & g(2) &= 0.781302412896486\ldots
\label{eq:lgconstants}
\end{align}
where $\lambda$ is the Catalan constant. Note that by using the general symmetry property obeyed by all absolutely convergent $\sigma_n^{(m)}$, given in \eqref{eq:invtau}, values for opposite aspect ratios can be obtained.

\subsection{Angle dependent orders ($m \neq 0$)}
Expressions for $S_{l,m,n}(u;\tau,a)$, where $m$ is a non-zero even integer, can be obtained from a unified procedure. Divergent Eisenstein sums do not feature as in $m=0$, only conditionally convergent sums, which can all be treated using the results of Section \ref{sec:nonmod}. Again, expressions for square lattices were derived by Nicorovici et al.\ using the Poisson summation formula,\autocite{nicorovici1996analytical} and its generalization to arbitrary lattices is stated briefly. This begins with
\begin{equation}
\sum_h \hat{f}(\bv{K}_h) = \frac{A}{(2\pi)^2} \sum_p f(\bv{R}_p),
\end{equation}
where the Fourier transform is defined
\begin{equation}
f(\bv{R}_p) = \int \hat{f}(\bv{K}) e^{-i\bv{K}\cdot\bv{R}_p} d\bv{K}.
\end{equation}
Expanding the exponential using the Jacobi--Anger identity and inserting the summand of \eqref{eq:Sdef} as $\hat{f}(\bv{K})$ yields
\begin{equation}
f(\bv{R}_p) = 2\pi e^{im\varphi_p} (-i)^m \int_0^\infty \frac{1}{K^{n-1}} J_l(Ku) J_m(KR_p) dK,
\end{equation}
simplified using the orthogonality of $e^{im\psi}$ angular terms. The integral is a Weber--Schafheitlin integral, which has a result in terms of the hypergeometric series\autocite{abramowitz1964handbook}
\begin{equation}
\frac{u^l (m-t-1)!}{2^{n-1} R_p^{m-2t} l!t!} {}_2F_1\left(-t, m-t; l+1; \left(\frac{u}{R_p}\right)^2\right),
\end{equation}
introducing the variable $-2t = l-m-n+2$. For non-negative $t$, the hypergeometric series terminates, generating a polynomial of order $t$, to yield
\begin{equation}
\begin{aligned}
S_{l,m,n}(u;\tau,a) &= \frac{A}{2\pi} (-i)^m \sum_p e^{im\theta_p} \frac{1}{2^{n-1}} \sum_{s=0}^t (-1)^s \frac{(m-t+s-1)!}{s!(t-s)!(l+s)!} \frac{u^{2s+l}}{R_p^{m-2t+2s}}\\
&= \frac{A}{\pi} \frac{(-i)^m}{2^n} \sum_{s=0}^t (-1)^s \frac{(m-t+s-1)!}{s!(t-s)!(l+s)!} \frac{u^{2s+l}}{a^{m-2t+2s}} \sigma_{m-2t+2s}^{(m)}(\tau).
\end{aligned}
\label{eq:poissonweber}
\end{equation}

The formula \eqref{eq:poissonweber} does not hold for all values of ($l$, $m$, $n$), and fails for negative $t$ for example. Nevertheless, all orders can be generated beginning with $S_{2,m,2}(u;\tau,a)$ and exploiting the recurrence relations \eqref{eq:recur1}--\eqref{eq:recur4}. Thus, we define
\begin{equation}
S_{2,m,2}(u;\tau,a) = \frac{i^m A}{4\pi} \sum_{k=0}^{m/2-1} D_k(m) \frac{u^{2k+2}}{a^{2k+2}} \sigma_{2k+2}^{(m)}(\tau),
\end{equation}
where
\begin{equation}
D_k(m) = (-1)^k \frac{(\frac{m}{2}+k)!}{k!(k+2)!(\frac{m}{2}-k-1)!}.
\label{eq:Dcoeff}
\end{equation}
For $m=2,4$, this evaluates to
\begin{align}
S_{2,2,2}(u;\tau,a) &= -\frac{A}{8\pi}u^2\frac{\sigma_2^{(2)}(\tau)}{a^2}, & S_{2,4,2}(u;\tau,a) &= \frac{A}{4\pi}\left[u^2\frac{\sigma_2^{(4)}(\tau)}{a^2}-u^4\frac{\sigma_4^{(4)}(\tau)}{a^4}\right],
\end{align}
while for $m=6$,
\begin{equation}
S_{2,6,2}(u;\tau,a) = -\frac{A}{8\pi}\left[3u^2\frac{\sigma_2^{(6)}(\tau)}{a^2} - 8u^4\frac{\sigma_4^{(6)}(\tau)}{a^4} + 5u^6\frac{\sigma_6^{(6)}(\tau)}{a^6}\right].
\end{equation}

One application of recurrence relation \eqref{eq:recur4} yields, for example,
\begin{equation}
S_{1,4,3}(u;\tau,a) = \left(\frac{A}{2\pi}\right)^2 \frac{u}{2} \frac{\sigma_2^{(4)}(\tau)}{a^2} + \frac{A}{4\pi}\left[-\frac{u^3}{2} \frac{\sigma_2^{(4)}(\tau)}{a^2} + \frac{u^5}{4} \frac{\sigma_4^{(4)}(\tau)}{a^4}\right],
\end{equation}
and subsequently applying \eqref{eq:recur3} yields
\begin{equation}
S_{2,4,4}(u;\tau,a) = \left(\frac{A}{2\pi}\right)^2 \frac{u^2}{8} \frac{\sigma_2^{(4)}(\tau)}{a^2} + \frac{A}{4\pi}\left[-\frac{u^4}{12} \frac{\sigma_2^{(4)}(\tau)}{a^2} + \frac{u^6}{32} \frac{\sigma_4^{(4)}(\tau)}{a^4}\right].
\end{equation}
Successive application of the recurrence relations produces the general form
\begin{equation}
\begin{split}
S_{l,m,n}(u;\tau,a) &= i^m\sum_{k=1}^{(n-l)/2} \left(\frac{A}{2\pi}\right)^{2k} B_{2k}(l,n) u^{n-2k} \frac{\sigma_{2k}^{(m)}(\tau)}{a^{2k}}\\ 
&+ \frac{i^mA}{4\pi}\sum_{k=\max\{0,(l-n)/2\}}^{m/2-1} C_{2k}(l,n) D_k(m) u^{n+2k} \frac{\sigma_{2k+2}^{(m)}(\tau)}{a^{2k+2}},
\label{eq:Sm}
\end{split}
\end{equation}
where the coefficients $B_{2k}(l,n)$, $C_{2k}(l,n)$, and $D_k(m)$ are given in \eqref{eq:Bcoeff}, \eqref{eq:Ccoeff}, and \eqref{eq:Dcoeff} respectively. The lower limit of the second sum is $(l-n)/2$, but cannot be negative. Again, if $(n-l)/2 < 1$, the first sum does not contribute, while the second sum also ceases to contribute if $(l-n)/2 > m/2 - 1$ and $S_{l,m,n}(u;\tau,a)$ is identically zero.

The most commonly encountered orders of $S_{l,m,n}(u;\tau,a)$ are when $n-l\leqslant m$. Then, \eqref{eq:Sm} only contains orders of $\sigma_n^{(m)}$ with $m \geqslant n$. The efficient methods of Section \ref{sec:fourier} may then be used to evaluate \eqref{eq:Sm}, specifically \eqref{eq:sigmanm}. Furthermore, for the square, hexagonal, ($2:1$) and ($\sqrt{3}:1$) rectangular lattices, the exact results of Table \ref{tab:sigma} apply, leading to entirely closed-form expressions for $S_{l,m,n}(u;\tau,a)$. When $n-l>m$, sums $\sigma_n^{(m)}(\tau)$ of orders $m<n$ begin to appear, for which numerical results can be generated by \eqref{eq:hyper}. Note that even though $S_{l,m,n}(u;\tau,a)$ are absolutely convergent over the reciprocal lattice, they are expressed in terms of $\sigma_2^{(m)}(\tau)$ which are conditionally convergent, which contain non-physical extraordinary contributions. If instead the regularized $\tilde{\sigma}_2^{(m)}(\tau)$ are used, derived using \eqref{eq:nonmod}, correct results for $S_{l,m,n}(u;\tau,a)$ are obtained.

\section{Evaluation over displaced lattices}
\label{sec:displaced}
In Section \ref{sec:gamma}, results were presented for the Bessel-modulated lattice sums
\begin{equation}
S_{l,m,n}(u;\tau,a) = \sideset{}{'}\sum_h \frac{J_l(K_hu)}{K_h^n}e^{im\psi_h},
\label{eq:Sdefrepeated}
\end{equation}
and were evaluated over two-dimensional lattices $\overbar{\Omega}$ centered on the origin in reciprocal space, such that each lattice point corresponds to the center of a different Brillouin zone. However, $S_{l,m,n}$ can be evaluated over any lattice, and instead of considering the lattice of all $\Gamma$ points, we now evaluate over the lattice of all $M$ points, for example. This can be achieved in several ways; the first is to replace $\bv{K}_h$ with $\bv{K}_h + \bv{k}_0$ and $\psi_h$ with $\mathrm{arg}(\bv{K}_h + \bv{k}_0)$ in \eqref{eq:Sdefrepeated}, where $\bv{k}_0$ is a vector which denotes a high symmetry point in the first Brillouin zone. Equivalently, we can leave the summand unchanged and modify the underlying point set $\overbar{\Omega}$, as considered here. 

\begin{figure}[t]
\centering
\subfloat[\label{fig:recipfig}]{%
\includegraphics[width=0.45\textwidth]{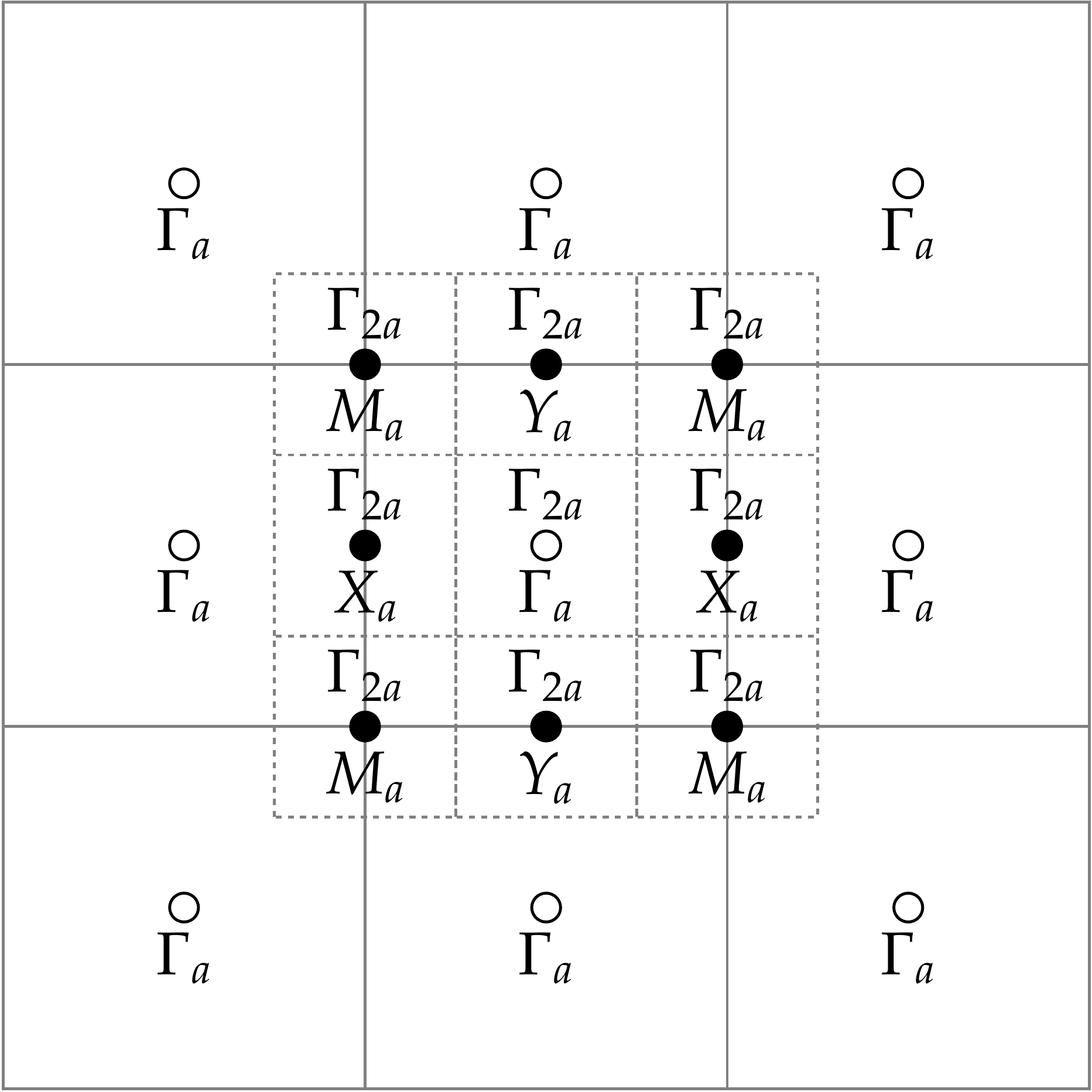}
}\quad\quad
\subfloat[\label{fig:hexfig}]{%
\includegraphics[width=0.39\textwidth]{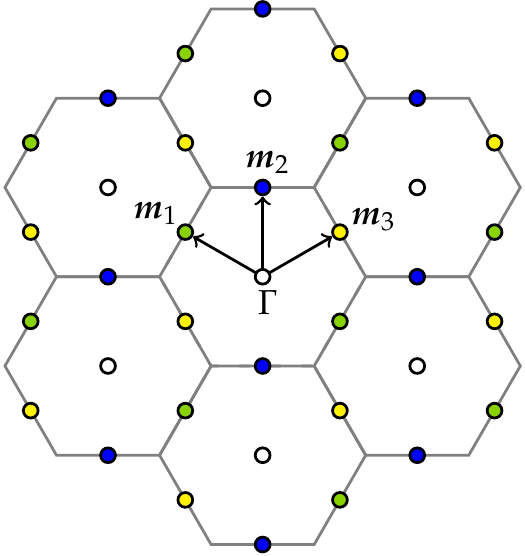}
}
\caption{ \protect\subref{fig:recipfig} Set of points for square lattice of period $a$ (open circles; $\Gamma_a$) and period $2a$ (filled circles; $\Gamma_{2a}$), in reciprocal space with Brillouin zones overlaid. The $\Gamma_{2a}$ set comprises the union of all $\Gamma$, $X$, $Y$, and $M$ points for a lattice of period $a$ (denoted with subscript $a$). Additional filled circles are omitted in neighboring cells. \protect\subref{fig:hexfig} Set of all $\Gamma$ (open circles) and $M$ points (colored circles) for hexagonal lattice, with Brillouin zones overlaid. The $M$ points are members of the lattice with twice the period and comprise the union of origin-centered hexagonal lattices translated by $\bv{m}_1$ (green), $\bv{m}_2$ (blue) and $\bv{m}_3$ (yellow).\label{fig:bothfig} 
}
\end{figure}

We proceed by first detaching from the typical interpretation of an origin-centered lattice as comprising the set of all $\Gamma$ points. This same set of points can also be regarded as the union of high symmetry points of a different lattice, for example, an origin-centered square lattice in reciprocal space is the union of all the $\Gamma$, $X$, $Y$, and $M$ points of a square lattice with double the period, as shown in Figure \ref{fig:bothfig}\subref{fig:recipfig}. Another example is that the reciprocal lattice of a rectangular array $\overbar{\Omega}(i/2,2a)$ comprises the $\Gamma$ and $X$ points of the square lattice $\overbar{\Omega}(i,a)$. Thus, it is possible to construct an array comprising just the $X$, $Y$, or $M$ points of a square reciprocal lattice using various origin-centered square and rectangular lattices.

Using the principles described, we obtain multi-set expressions to enable the evaluation of displaced lattice sums in terms of the sums already evaluated in Section \ref{sec:gamma}. Key to their validity is the regularization of divergent and conditionally convergent terms. We present expressions for all high symmetry point displacements of square and hexagonal lattices, although this method can be extended to consider other Bravais lattices. We also apply this method to sums over the real lattice, such as the generalized Eisenstein series $\sigma_n^{(m)}$, and so all of the expressions in this section which introduce an offset to the reciprocal lattice can be used to introduce an offset to the direct lattice instead.

\subsection{Square lattices} 
\label{sec:Slmnsquaredisp}
For a square lattice, the set of all $X$ points can be explicitly written as\begin{equation}
\overbar{\Omega}^X(i,a) = \left\{ \left. 2\pi\left(\frac{h_1}{a},\frac{h_2}{a}\right) + \left(\frac{\pi}{a},0\right) \right|h_1,h_2 \in \mathbb{Z} \right\}.
\label{eq:overbarXlong}
\end{equation}
The above set is contained within $\overbar{\Omega}(i/2,2a)$, and so \eqref{eq:overbarXlong} can be represented by simply subtracting extraneous points corresponding to the original undisplaced points via
 \begin{equation}
\overbar{\Omega}^X(i,a) = \overbar{\Omega}( i/2,2a) - \overbar{\Omega}(i,a).
\label{eq:squaresetX}
\end{equation}
Here, the set of points on the left-hand side is equal to the addition and subtraction, counting multiplicity, of sets of points on the right-hand side. A similar treatment for the $Y$ point set gives
\begin{equation}
\overbar{\Omega}^Y(i,a) =\overbar{\Omega}( 2i,a) - \overbar{\Omega}(i,a).
\label{eq:squaresetY}
\end{equation}
As mentioned, the $M$ points of a square lattice of period $a$ are contained in the $\Gamma$ points of a square lattice of period $2a$. The set of extraneous points are now all the $X$, $Y$, and the original $\Gamma$ points, which may by subtracted through the relation
\begin{equation}
\begin{split}
\overbar{\Omega}^M(i,a) &= \overbar{\Omega}(i,2a) - \overbar{\Omega}^X(i,a) - \overbar{\Omega}^Y(i,a) - \overbar{\Omega}(i,a)\\ 
&= \overbar{\Omega}(i, 2a) - \overbar{\Omega}(i/2,2a) + \overbar{\Omega}(i,a) - \overbar{\Omega}(2i,a).
\end{split}
\label{eq:squaresetM}
\end{equation}
To demonstrate the use of these identities, we present the general form for $S_{l,4,n}$ over $\overbar{\Omega}^M(i,a)$ as given by 
 \begin{subequations}
\begin{equation}
\begin{split}
S^M_{l,4,n}(u ) = \sum_{k=1}^{(n-l)/2} B_{2k}(l,n) u^{n-2k} \left(\frac{a}{ 2 \pi}\right)^{2k} \left[\left(2^{2k} + 1\right) \sigma_{2k}^{(4)}(i) - \left( 2^{2k} + 2^{-2k} \right) \sigma_{2k}^{(4)}\left( 2i \right)\right] \\ 
+ \frac{1}{4\pi} \sum_{k=\mathrm{max}\left\{0,(l-n)/2\right\}}^{1} C_{2k}(l,n) D_k(4) u^{n+2k} \frac{1}{(2a)^{2k}} \left[ 2 \sigma_{2
k+2}^{(4)}(i) - \left( 2^{1+2k} + 2^{-1} \right) \sigma_{2k+2}^{(4)}\left(2i \right)\right],
\end{split}
\label{eq:S4Mgen}
\end{equation}
which follows from \eqref{eq:Sm}, \eqref{eq:squaresetM}, and where the regularized forms for $\sigma_n^{(m)}$ must be used. For a particular $(l,n)$ pair we obtain
\begin{equation} 
S_{1,4,5}^M(u) =\Gamma \! \left(\tfrac{1}{4}\right)^8 \left[-\frac{a^4 u}{3\cdot 2^{11} \pi ^6}+\frac{a^2 u^3}{2^{13} \pi ^5}-\frac{u^5}{9\cdot 2^{12} \pi ^4}+\frac{u^7}{15 \cdot 2^{15} a^2 \pi ^3} \right],
\end{equation}
\end{subequations}
after using the closed-form expressions for $\sigma_{2,4}^{(4)}$ on Table \ref{tab:sigma} as well as the identities \eqref{eq:invtau} and \eqref{eq:invtau2m}.
Finally, we note that $S_{l,m,n}$ is vanishing when $m$ is not an integer multiple of $4$ and is evaluated over lattices with 4-fold symmetry, such as origin-centered sets or over the set of all $M$ points. In these instances, $S^{X,Y}_{l,m,n}$ is identical to a rectangular lattice sum, following \eqref{eq:squaresetX} and \eqref{eq:squaresetY}.

\subsection{Hexagonal lattices} \label{sec:Slmnhexdisp}
For the hexagonal lattice, the treatment is not as straightforward. In the square lattice, there are just as many $\Gamma$ points as $X$, $Y$, or $M$ points, across the whole reciprocal space. However for the hexagonal lattice there are 3 times as many $M$ points, and twice as many $K$ points, as $\Gamma$ points. This is shown in Figure \ref{fig:bothfig}\subref{fig:hexfig} where there are three different subsets of $M$ points that can each be obtained by three different translations of $\Gamma$ points. These translations are given by $ \bv{m}_{1} = \left( -\tfrac{\pi}{a}, \tfrac{ \pi}{\sqrt{3} a} \right)$, $\bv{m}_2 = \left( 0, \tfrac{2 \pi}{\sqrt{3} a} \right)$, and $\bv{m}_3 = \left( \tfrac{\pi}{a}, \tfrac{ \pi}{\sqrt{3} a} \right)$ and their respective subsets are
\begin{subequations}
\begin{align}
\overline{\Omega}^{{M}_{j}}(e^{i \pi/3},a) &= \left\{ \left. \frac{2 \pi}{\sqrt{3} a} \left( \sqrt{3} h_1,2h_2-h_1 \right) + \bv{m}_{j} \right| h_1,h_2 \in \mathbb{Z} \right\}.
\label{eq:omegaMHex}
\end{align}
The set of all $M$ points is the union of all these subsets
\begin{align}
\overline{\Omega}^{M}(e^{i \pi /3},a) &= \overline{\Omega}^{M_1}(e^{i \pi /3},a) + \overline{\Omega}^{M_2}(e^{i \pi /3},a) + \overline{\Omega}^{M_3}(e^{i \pi /3},a).
\label{eq:Munion}
\end{align}
\end{subequations}
Similarly, for the $K$ point set we use the translation vectors $\bv{k}_1 = \left( -\frac{2 \pi}{3 a},\frac{2 \pi}{\sqrt{3} a} \right)$ and $\bv{k}_2 = \left( \frac{2 \pi}{3 a},\frac{2 \pi}{\sqrt{3} a} \right)$ and define the sets
 \begin{subequations}
\begin{align}
\overline{\Omega}^{K_j}(e^{i \pi /3},a) &= \left\{ \left. \frac{2 \pi}{\sqrt{3} a} \left( \sqrt{3} h_1,2h_2-h_1 \right) + \bv{k}_{j} \right| h_1,h_2 \in \mathbb{Z} \right\}, \\ \label{eq:Kunionset}
\overline{\Omega}^{K}(e^{i \pi /3},a)&= \overline{\Omega}^{K_1}(e^{i \pi /3},a) + \overline{\Omega}^{K_2}(e^{i \pi /3},a). 
\end{align}
 \end{subequations}
Since the set of $M$ points are contained in a lattice with twice the lattice constant, we may apply the same geometric principles as for the square lattice to obtain
\begin{equation}
\overbar{\Omega}^M(e^{i \pi /3},a) = \overbar{\Omega}(e^{i \pi /3},2a) - \overbar{\Omega}(e^{i \pi /3},a).
\label{eq:hexsetM}
\end{equation}
Meanwhile, the set of all $\Gamma$ and $K$ points only coincide if the $\Gamma$ points are rotated and scaled 
\begin{equation}
\overbar{\Omega}^K(e^{i \pi /3},a)= C_4\overbar{\Omega}(e^{i \pi /3},\sqrt{3}a) - \overbar{\Omega}(e^{i \pi /3},a),
\label{eq:hexsetK}
\end{equation}
where $C_4$ denotes rotation of the set of points by $\pi/2$. Note that this requires lifting the restriction imposed on \eqref{eq:Sdefrepeated}, of orienting $\bv{\hat{e}}_1$ along the $x$-axis. However, its effect on \eqref{eq:Sdefrepeated} can be determined by the replacement $\psi_h \rightarrow \psi_h + \pi/2$, so all terms of the lattice attract a global phase, with $S_{l,m,n} \rightarrow \mathrm{e}^{im\pi/2} S_{l,m,n}$.

When the order $m$ of the sum is an integer multiple of 6, the sum evaluated over each of the subsets in \eqref{eq:omegaMHex} is identical, so \eqref{eq:hexsetM} can be used to obtain each of the identical terms on the right-hand side of \eqref{eq:Munion}, and similarly for \eqref{eq:Kunionset}. If $m$ is not a multiple of $6$, then this symmetry cannot be exploited, so explicit expressions are required for the subsets
\begin{equation}
\label{eq:Mthree}
\begin{split}
&\overline{\Omega}^{{M}_{2}}(e^{i \pi/3},a) = \overbar{\Omega}(\sqrt{3}i,a) - \overbar{\Omega}(e^{i \pi/3},a),\\
\overline{\Omega}^{{M}_{1}}(e^{i \pi/3},a) = &C_3 \overline{\Omega}^{{M}_{2}}(e^{i \pi/3},a), \quad \overline{\Omega}^{{M}_{3}}(e^{i \pi/3},a) = C_{-3} \overline{\Omega}^{{M}_{2}}(e^{i \pi/3},a) ,
 \end{split}
\end{equation}
where $C_{\pm 3}$ denotes a rotation by $\pm \pi/3$, meaning that the corresponding $S_{l,m,n}$ term collects a phase factor of $\mathrm{e}^{ \pm i m \pi / 3 }$. Meanwhile, the $\overline{\Omega}^{{K}_{1}}(e^{i \pi/3},a)$ and $\overline{\Omega}^{{K}_{2}}(e^{i \pi/3},a)$ subsets cannot be individually evaluated using origin-centered sums because they lack the requisite 2-fold symmetry about the origin, and so do not qualify as origin-centered lattices.

An example of $S_{l,m,n}$ over the $M$ points of a hexagonal array is the general form for $m=0$, given by
\begin{subequations}
\begin{equation}
\begin{split}
S_{l,0,n}^M(u) &= \sum_{k=2}^{(n-l)/2} \left(\frac{1}{2\pi}\right)^{2k} B_{2k}(l,n) u^{n-2k}\left({\sqrt{3}a} \right)^{2k} \left( 1 - \frac{1}{2^{2k}} \right)\sigma_{2k}^{(0)}\! \! \left(e^{i \pi/3}\right) \\
 &- \frac{ 3\sqrt{3}a^2 }{4\pi} B_2(l,n) u^{n-2} \left[ \log\left( \frac{u}{2^{4/3}} \right) + C_l(l,n) + \log\left( \frac{3^{1/4} \Gamma(\frac{1}{3})^3 }{2 \pi a}\right)\right],
\end{split}
\end{equation}
which for a particular $(l,n)$ pair takes the form
\begin{equation} %this has been validated numerically
 S_{2,0,6}^M(u) = \frac{135 a^4 u^2 g(2)}{2048 \pi^2} - \frac{17 a^2 u^4}{512 \sqrt{3} \pi} + \frac{\sqrt{3} a^2 u^4 }{128\pi} \log\left(\frac{u}{2^{4/3}}\right) + \frac{3a^2 u^4}{128 \sqrt{3}\pi} \log\left(\frac{3^{1/4} \Gamma(\tfrac{1}{3})^3}{2\pi a}\right),
\end{equation}
\end{subequations}
where $g(2)$ is defined in \eqref{eq:lgconstants}.

\subsection{Evaluation of $\sigma_{n}^{(m)}(\tau)$ over displaced point sets} 
\label{sec:sigmanmsquaredisp}
\begin{figure}[t]
\centering
\includegraphics[width=0.45\textwidth]{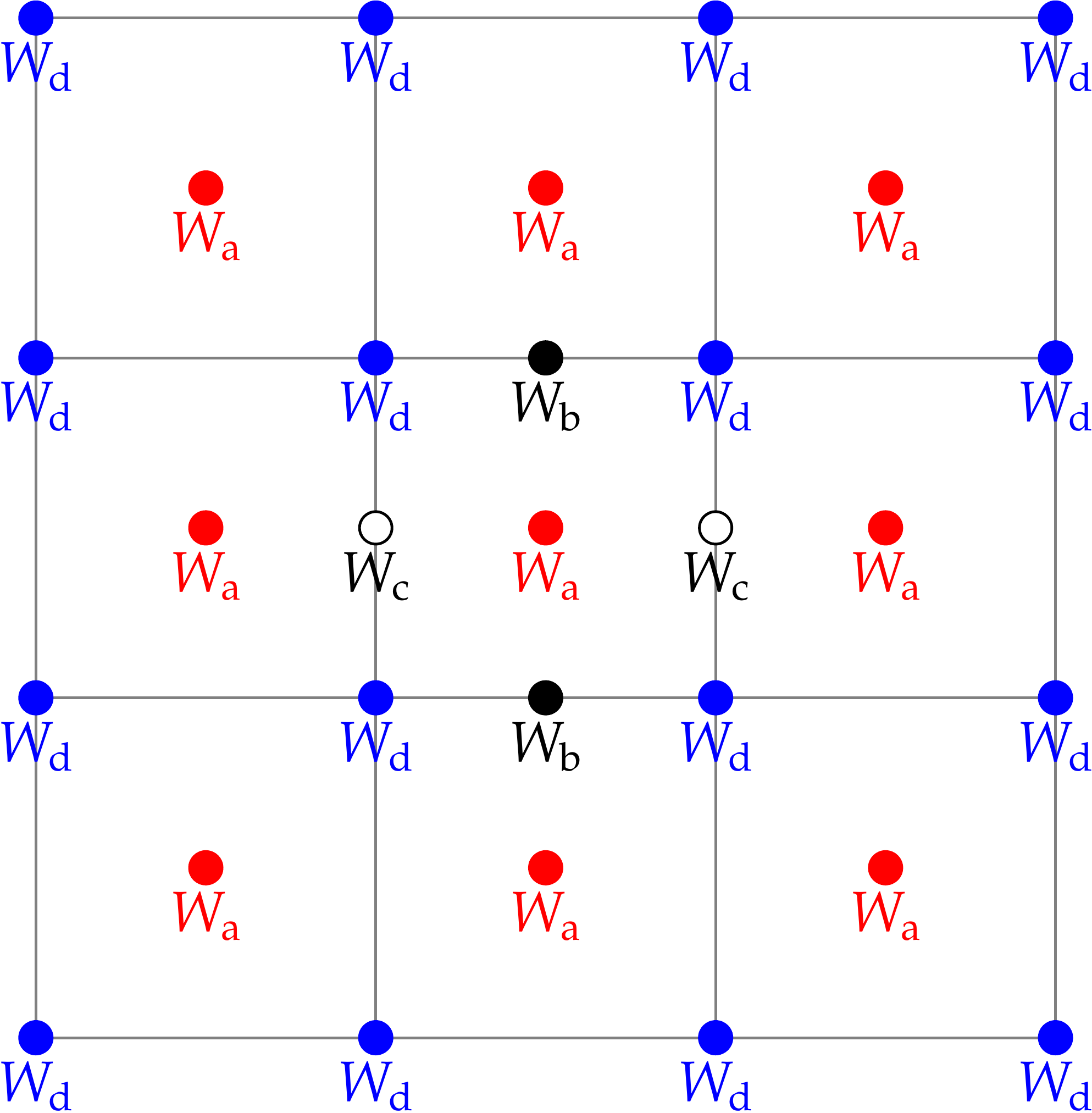}
\caption{An outline of the square two-dimensional direct lattice $\Omega(i,a)$ comprising the origin-centered coordinates $W_\mathrm{a}$ (red circles), where solid lines denote the edges of each unit cell. Also shown is the set of all Wyckoff positions $W_{\mathrm{d}}$ (blue circles), which is equivalent to a translation of the set of $W_\mathrm{a}$ points by $(a/2,a/2)$. The high symmetry points $W_\mathrm{b}$ (solid circles) and $W_\mathrm{c}$ (open circles) are also indicated at the edge of the fundamental cell. \label{fig:realfig}}
\end{figure}

Results for the generalized Eisenstein series
\begin{equation}
\label{eq:sigmanmcopy}
\sigma_n^{(m)}(\tau) = \sideset{}{'}\sum_{p } \frac{e^{-im\varphi_p}}{R^n_p},
\end{equation}
were presented in Section \ref{sec:eisenstein} for a two-dimensional real lattice $\Omega$ centered about the origin. Following \eqref{eq:Sdefrepeated}, we relax the restriction that \eqref{eq:sigmanmcopy} be evaluated over an origin-centered lattice and consider sets of Wyckoff positions over all direct lattice cells. Wyckoff positions are points of high symmetry in the direct lattice, corresponding to coordinates in the unit cell which possess a multiplicity and symmetry, and specify where additional coordinates must be located in the unit cell so that the symmetry of the lattice is preserved. Figure \ref{fig:realfig} shows the four Wyckoff positions for a square lattice, denoted space group $p2$: $W_\mathrm{a} = (0,0)$, $W_\mathrm{b} = (0,a/2)$, $W_\mathrm{c} = (a/2,0)$, and $W_d = (a/2,a/2)$, where $a$ is the period. We remark that the lattice given by the union of the $\Omega^{W_\mathrm{a}}(i,a) $ and $\Omega^{W_\mathrm{d}}(i,a) $ lattices constitutes a diatomic lattice, enabling lattice sums to be evaluated over each of the two constituents separately.

To demonstrate that the multi-set identities for the reciprocal lattice also extend to the real lattice, we follow the procedure outlined in Section \ref{sec:Slmnsquaredisp} and present the corresponding expression to \eqref{eq:squaresetM}, which takes the form 
\begin{equation}
\begin{split}
\Omega^{W_\mathrm{d}}(i,a) &= \left\{\left. a (p_1,p_2) + \left(\frac{a}{2},\frac{a}{2}\right)\right| p_1,p_2 \in \mathbb{Z} \right\} \\
&=\Omega(i,a/2) + \Omega(i,a) - \Omega(i/2,a) - \Omega(2i,a/2),
\end{split}
\label{eq:Wbreal}
\end{equation}
where in contrast to the reciprocal lattice, the set of $W_\mathrm{a}$, $W_\mathrm{b}$, and $W_\mathrm{c}$ points of a square lattice are contained in the square direct lattice of half the period, instead of double the period of the reciprocal lattice. Using the values of Table \ref{tab:sigma} and the multi-set identity \eqref{eq:Wbreal} we obtain the two following closed-form representations
\begin{equation}
\sigma_{4,W_\mathrm{d}}^4(i) = -\frac{\Gamma(\tfrac{1}{4})^8}{192\pi^2}, \qquad \sigma^4_{2,W_\mathrm{d}}(i) = -\frac{\Gamma(\tfrac{1}{4})^8}{128\pi^3},
\end{equation}
where the expression for $\sigma^4_{2,W_\mathrm{d}}(i)$ is significant because it is conditionally convergent and cannot be obtained by direct summation. Here, the use of regularized sums is crucial to the success of \eqref{eq:Wbreal}.

\section{Summary}
We treat Eisenstein series in Section \ref{sec:eisenstein}, obtaining a general formula \eqref{eq:hyper} to evaluate generalized Eisenstein series $\sigma_n^{(m)}$, applicable to all lattices and all even orders $m$ with $n \geqslant 2$. The result is given as a Fourier series, and exhibits rapid numerical convergence. In the special case of $m \geqslant n$, this formula converts generalized Eisenstein series into derivatives of conventional Eisenstein series, \eqref{eq:sigmanm}, which may subsequently be converted entirely into products of conventional Eisenstein series $G_k(\tau)$ using Ramanujan's derivative identities \eqref{eq:eisenderiv} and the recursion relation \eqref{eq:eisenrecur}. This permits closed-form evaluations for all orders $m \geqslant n$ over many high symmetry lattices via the many known special values of the Dedekind $\eta$-function, in a procedure described in Section \ref{sec:exact}. We perform this procedure to obtain closed-form results for several important lattices, and these are displayed on Table \ref{tab:sigma}.

For conditionally convergent orders $\sigma_2^{(m)}$, all of our evaluation methods yield results which may be regularized to give physically meaningful results. The summation order we impose conforms to the Eisenstein summation method, which is known to include a non-physical contribution. In Section \ref{sec:nonmod}, we identify this component by comparing $\sigma_2^{(m)}$ to an absolutely convergent sum that obeys the necessary geometric identities, analytically continued to the conditionally convergent case. This yields a simple formula \eqref{eq:nonmod} which allows the non-physical contribution to be subtracted, thereby regularizing all results conforming to the Eisenstein summation order, including \eqref{eq:hyper}, \eqref{eq:sigmanm}, and Table \ref{tab:sigma}.

In Section \ref{sec:gamma}, we apply these results for generalized Eisenstein series to evaluate the cylindrical harmonic sums $S_{l,m,n}$, the two being related by Poisson summation formula. For orders with angular variation ($m \neq 0$), we obtain a result applicable to all valid orders, \eqref{eq:Sm}. This demonstrates the success of the regularization procedures, as $S_{l,m,n}$ have unambiguous values by virtue of their absolute convergence, which can be validated by direct summation, yet $S_{l,m,n}$ are expressed in terms of conditionally convergent Eisenstein series. The angular invariant $S_{l,0,n}$ feature a divergent sum, which is regularized using Kronecker's second limit formula. The general result for orders $l<n$ is given by \eqref{eq:S0upper}. For the remaining orders, two special forms exist due to the regularization of the divergent sum, \eqref{eq:S0diag} and \eqref{eq:S0lower}.
 
Finally, in Section \ref{sec:displaced} we derive identities that express $\sigma_n^{(m)}$ and $S_{l,m,n}$ sums over displaced lattices entirely in terms of the origin-centered lattices of Sections \ref{sec:eisenstein} and \ref{sec:gamma}. We consider high symmetry displacements of square reciprocal lattices in Section \ref{sec:Slmnsquaredisp}, presenting identities for the lattices of all $X$ points \eqref{eq:squaresetX}, $Y$ points \eqref{eq:squaresetY}, and $M$ points \eqref{eq:squaresetM}. The hexagonal lattice is treated in Section \ref{sec:Slmnhexdisp}, with identities for the $K$ points \eqref{eq:hexsetK} and $M$ points \eqref{eq:hexsetM}. The set of all hexagonal $M$ points can be further decomposed into its constituent sublattices \eqref{eq:Mthree}, but not the $K$ points since the sublattices lack sufficient symmetry. To demonstrate our method applies to direct lattices as well as reciprocal lattices we also present an identity for the direct square lattice in \eqref{eq:Wbreal}.

\section*{Postscript}
A very recent paper on arXiv (1606.04355) by Yakubovich, Drygas, and Mityushev contains some results and methods complementary to those given here.\autocite{yakubovich2016closed}

\printbibliography
\end{document}